\documentclass[]{interact}
\usepackage{lmodern}

\usepackage{epstopdf}

\usepackage[numbers,sort&compress]{natbib}%
\bibpunct[, ]{[}{]}{,}{n}{,}{,}%
\makeatletter%
\def\NAT@def@citea{\def\@citea{\NAT@separator}}%
\makeatother%

\usepackage{wrapfig}
\usepackage[utf8]{inputenc} 

\usepackage[T1]{fontenc}    
\usepackage{hyperref}       
\usepackage{hyperref}       

\usepackage[caption=false]{subfig}

\usepackage{url}            
\usepackage{booktabs}       
\usepackage{amsfonts}       
\usepackage{nicefrac}       
\usepackage{microtype}      
\usepackage{xcolor}         
\usepackage{graphicx}       
\usepackage{amsmath}        
\usepackage{amssymb}        
\usepackage{multirow}
\usepackage{mdframed}
\usepackage{ascmac}
\usepackage[most]{tcolorbox}

\usepackage{amsthm}

\theoremstyle{plain}%

\theoremstyle{definition}

\theoremstyle{remark}

\usepackage{amssymb}
\usepackage{mathtools} 

\usepackage[final,commandnameprefix=ifneeded]{changes}
\setaddedmarkup{\textcolor{red}{#1}}
\setdeletedmarkup{\textcolor{red}{\sout{#1}}}

\title{Co-Creative Learning via Metropolis-Hastings Interaction between Humans and AI}

\author{
\name{Ryota Okumura\textsuperscript{a,\textdagger}, Tadahiro Taniguchi\textsuperscript{a,b,\textdagger,*}, Akira Taniguchi\textsuperscript{a}, and Yoshinobu Hagiwara\textsuperscript{a,c}}
\affil{\textsuperscript{a}College of Information Science and Engineering, Ritsumeikan University, Shiga, Japan; \textsuperscript{b}Graduate School of Informatics, Kyoto University, Kyoto, Japan; \textsuperscript{c}Faculty of Science and Engineering, Soka University, Tokyo, Japan}
}

\begin{document}

\maketitle
\begingroup
\renewcommand{\thefootnote}{\textdagger}
\footnotetext{Equal contribution.}
\renewcommand{\thefootnote}{*}
\footnotetext{CONTACT Tadahiro Taniguchi. Email: taniguchi@i.kyoto-u.ac.jp}
\endgroup

\begin{abstract}
We propose co-creative learning as a novel paradigm where biological and artificial agents (e.g., humans and robots) mutually integrate their partial perceptual information and knowledge to construct shared external representations, a process we interpret as symbol emergence. Unlike traditional AI teaching based on unilateral knowledge transfer, this addresses the challenge of integrating information from inherently different modalities. We empirically test this framework using a human-AI interaction model based on the Metropolis-Hastings naming game (MHNG), a decentralized Bayesian inference mechanism. In an online experiment, $69$ participants played a joint attention naming game (JA-NG) with one of three computer agent types (MH-based, always-accept, or always-reject) under partial observability. Results show that human-AI pairs improved their categorization accuracy through interaction, and that an MH-based agent converged on a shared sign system aligned with the two agents' joint observations significantly more closely than an agent learning from its own observations alone, a process underpinned by the collaborative minimization of their collective free energy. No statistically significant difference between the MH and Always Accept conditions was detected after Holm correction. Unlike unconditional acceptance under Always Accept, however, the MH-based agent uses an evidence-sensitive acceptance rule based on its local generative model. Furthermore, human acceptance behavior aligned closely with the MH-derived acceptance probability. These findings provide the first empirical evidence for co-creative learning emerging in human-AI dyads via MHNG-based interaction. This suggests a promising path toward symbiotic AI systems that learn with humans, rather than from them, by dynamically aligning perceptual experiences, opening a new venue for symbiotic and organic AI alignment.
\end{abstract}


\section{Introduction}
\label{sec:introduction}

A fundamental challenge in robotics and artificial intelligence (AI) is how humans and autonomous robots can integrate their partial, heterogeneous information to achieve a collective goal in a shared environment. 
Understanding the principles of such collaborative intelligence has profound implications for the design of intuitive human-robot interaction (HRI), from assistive systems in healthcare and daily living to collaborative robots in manufacturing. In the context of AI, this challenge manifests as the problem of AI alignment, which is crucial as we move towards a future of human-robot symbiosis.
When (embodied) AI agents begin to explore the world autonomously, should we enforce our existing knowledge and values as the sole ground truth? Evidence suggests that AI systems, such as large language models (LLMs), already surpass average human performance in various domains.  

Traditionally, alignment is often viewed as a unilateral process: transferring human ethics, values, and knowledge to the AI agent and constraining its behavior accordingly. 
However, this top-down approach is particularly insufficient for robots designed to learn and adapt alongside humans in dynamic, real-world settings.
As we already observe a bidirectional influence where AI systems like LLMs impact human language use and decision-making \citep{yakura2024empirical,brinkmann2023machine,shin2023superhuman,geng2024chatgpt,kobak2025delving}. This mutual influence is a natural consequence of interaction, rendering unilateral alignment either insufficient or impossible. Given this inherent bilaterality, a new machine learning theory is needed to guide the development of beneficial symbiotic systems: systems where humans and AI co-create knowledge and understanding. 
This concept, which we term \textbf{co-creative learning}, is vital for addressing the ethical, legal, and social implications (ELSI) of robotics, as it offers a framework for humans and robots to ``create knowledge together,'' which is central to the growing notion Human-AI symbiotic/organic or bilateral/bidirectional alignment~\citep{shen2024towards}. 

To build a theoretical foundation for such co-creative systems, we bridge machine learning with insights from cognitive robotics on how agents achieve shared understanding. This work is situated in the research lineage of symbol emergence in robotics, which investigates how robots can autonomously form grounded concepts and communication systems through sensorimotor experience~\cite{taniguchi2016symbol,taniguchi2019symbol,taniguchi2024collective}.
A promising theoretical foundation for this challenge can be found in the Free Energy Principle (FEP) and its process theory, active inference, which provides a unifying account of how biological agents, such as brains, maintain their organization by making sense of their environment \citep{friston2010free, hohwy2013predictive}. A key question, then, is how this principle extends from a single agent to a collective of interacting agents. The Collective Predictive Coding (CPC) hypothesis addresses this by proposing that a group minimizes its \textbf{collective free energy} not only through individual learning (i.e., neural plasticity) but also by creating and adapting a shared communication system (i.e., semiotic plasticity) \citep{taniguchi2024collective}. This provides a new, cognitively grounded perspective on a central challenge in the field of Emergent Communication (EmCom), or symbol emergence: how communication systems and shared symbols arise from local interactions in a multi-agent system \citep{foerster2016learning, lazaridou2022emergent, peters2025emergent}.

The mathematical formulation of this collective free energy, which decomposes into individual and collective terms, has been explicitly provided in the context of modeling scientific activities \citep{taniguchi2025collective}. This perspective aligns with a broader line of inquiry within the active inference community that explores how agents collectively build shared structures, such as cognitive niches \citep{Constant2018variational} and cultural representations \citep{Veissiere2020thinking}, through joint inference. Crucially, generative model-based approaches that implement CPC, such as the Metropolis-Hastings naming game (MHNG), facilitate this process through decentralized Bayesian inference, a mechanism for what is also termed federated inference \citep{Friston2024federated}, performed via local interactions without requiring agents to access each other's internal states or exchange gradients \citep{hagiwara2019symbol, taniguchi2023emergent, taniguchi2024collective}.

The MHNG, as an algorithmic realization of CPC principles, offers a compelling model for how agents can achieve symbol emergence. Given that it operates in a decentralized, black-box manner, it was hypothesized that humans could be integrated into such a system. Indeed, recent work in experimental semiotics provided the first evidence supporting this, showing that human-human dyads playing a Joint Attention Naming Game (JA-NG) exhibit dynamics consistent with MHNG theory \citep{okumura2023metropolishastings}. These findings in human-human contexts raise a critical  question: can this same mechanism facilitate a deeper process of knowledge creation, which we term \emph{co-creative learning}, between humans and robots? Specifically, this study examines whether a human-AI pair, under partial observation, can leverage MHNG dynamics to perform decentralized Bayesian inference and thereby integrate their disparate perceptual inputs into a shared, grounded understanding that neither could attain independently.

This paper pursues this hypothesis. We begin by formally proposing \emph{co-creative learning} as a paradigm where humans and AI achieve symbol emergence by mutually integrating their partial perceptual information to construct shared representations. We then test this formulation empirically through a case study involving a JA-NG where human-robot pairs categorize partially observable stimuli. Our findings demonstrate that MHNG-mediated interaction enables effective integration of heterogeneous perceptual information. Critically, this co-creative process, driven by the MH acceptance mechanism, allows human-AI dyads to converge on a shared, grounded sign system that integrates both agents' partial observations, reaching a level of agreement with the joint-observation structure that neither agent attains from its own partial observation alone.

The main contributions of this paper are twofold, aiming to establish and demonstrate a novel approach to human-AI collaboration:
\begin{itemize}
\item We formally propose \textbf{co-creative learning}, a novel paradigm where humans and AI collaboratively achieve symbol emergence by mutually integrating partial perceptual information to build shared representations. This offers a new perspective on human-AI alignment, moving beyond traditional unilateral approaches.
\item We introduce an experimental semiotics framework based on the MHNG and provide compelling empirical evidence that human-AI pairs can effectively engage in this co-creative learning. 
\end{itemize}
Our findings show that MHNG-mediated interaction enables the dyad to integrate complementary partial observations into a shared sign system whose agreement with the joint-observation posterior is higher than that obtained by the AI under self-directed learning. This information-completion result provides a concrete basis for investigating systems in which humans and robots construct shared representations through interaction.

The remainder of this paper is structured as follows. Section 2 reviews prior studies related to our work. Section 3 details the preliminaries of our framework, namely the Metropolis-Hastings Naming Game (MHNG) and the Interpersonal Gaussian Mixture (Inter-GM) model. In Section 4, we establish the conceptual foundation of co-creative learning, providing its formal definition based on a generative model and the principle of collective free energy minimization. We describe the design of our human-robot interaction experiment in Section 5 and report on the results in Section 6. Finally, we discuss the broader implications of our findings in Section 7 and conclude the paper in Section 8.

\section{Related Work}
\label{sec:related_work}
\subsection{EmCom and Symbol Emergence}
Research on EmCom explores how communication systems, including language-like symbol structures, can arise in multi-agent systems through interaction and coordination \citep{lazaridou2022emergent, peters2025emergent, brandizzi2023toward}. These studies range from discriminative models used in referential games \citep{lazaridou2017multi, foerster2016learning} to generative approaches grounded in naming games and population-based protocols \citep{hagiwara2019symbol, taniguchi2023emergent, ueda2024lewis}. While EmCom research has increasingly incorporated multi-agent reinforcement learning settings, it still primarily focuses on agent–agent interactions and synthetic symbol systems. In contrast, Symbol Emergence in Robotics emphasizes grounding and embodiment by investigating how autonomous agents construct internal representations through sensorimotor coupling with the physical world \citep{taniguchi2016symbol, taniguchi2019symbol, cangelosi2015developmental}.
Recently, the CPC hypothesis has been proposed to unify generative approaches to symbol emergence and bridge cognitive theory and embodied interaction \citep{taniguchi2024collective}. CPC extends predictive coding \citep{hohwy2013predictive} and the free energy principle \citep{friston2010free} to multi-agent systems, positing that shared symbol systems emerge as a consequence of minimizing prediction error across agents with heterogeneous sensorimotor experiences, mediated by decentralized inference mechanisms such as MHNG. However, despite its theoretical relevance, CPC's implications for human–AI co-creative learning and AI alignment remain underexplored. This work addresses this gap through theoretical formalization and empirical validation of its core mechanism in a human–machine collaborative setting.

\subsection{Experimental Semiotics}
Experimental semiotics employs laboratory studies to investigate the emergence of human communication systems \citep{galantucci2009experimental, galantucci2011experimental}, demonstrating human creation of novel systems \citep{galantucci2005experimental} and cultural transmission via iterated learning \citep{kirby2002emergence, kirby2008cumulative, cornish2013iterated}. This field incorporates artificial agents \citep{steels1999spontaneous, steels2015talking} and computational models, such as probabilistic inference (e.g., Markov Chain Monte Carlo (MCMC)-like dynamics in human category learning \citep{sanborn2007markov}), to understand underlying mechanisms. Directly relevant, Okumura et al.~\cite{okumura2023metropolishastings} found that human acceptance in a joint attention naming game (JA-NG) aligns with Metropolis-Hastings naming game (MHNG) probabilities, supporting MHNG as a model for human symbol emergence dynamics. Building on this, our work extends the MHNG framework to human-robot co-creative pairs, examining whether similar stochastic-inference principles drive alignment across these heterogeneous agents.

\subsection{Other Usages of the Term ``Co-creative Learning''}
The term ``co-creative learning'' is often used in Human-AI/robot collaboration to describe systems that assist or augment human creativity \citep{ali2021creativity, sandoval2022hrci, carayannis2021cocreation, zoelen2021team}. While these works highlight mutual adaptation and shared creativity, they tend to emphasize user experience over formal models. This differs from our formal definition presented in this paper, where we specifically define \emph{co-creative learning} as a computational process where humans and AI integrate their distinct perceptual information and knowledge through interaction, mediated by decentralized inference mechanisms like MHNG. This leads to the emergence of shared representations or symbols that neither could construct alone.
Our perspective aligns with recent studies on AI-enhanced and hybrid collective intelligence \citep{peeters2021hybrid,cui2024collective} and human-AI coevolution \citep{pedreschi2025human}, which stress the feedback-driven, emergent nature of hybrid intelligence, often highlighting the quality of interaction among its components. Unlike many HCI and LLM-based approaches, our framework provides a computational foundation for modeling how shared understanding arises from perceptual heterogeneity.

\subsection{Mutual Adaptation and Human-AI Teaming}
A substantial body of empirical work has examined human-autonomy teaming, in which humans and autonomous agents coordinate interdependently toward shared goals \citep{oneill2022hat}. Within HRI, mutual adaptation has been studied through models of human adaptation, role-alternating training that promotes shared mental models, and shared-control schemes that arbitrate between user input and robot policies \citep{nikolaidis2017mutual, nikolaidis2015crosstraining, dragan2013policy, muraoka2024goal}. In human-AI teaming, research has emphasized human mental models and the establishment of common ground \citep{bansal2019beyond, yao2026humanagent}. Recent work further moves beyond the teacher-learner asymmetry of interactive machine learning toward bidirectional alignment, including settings in which users refine their own concept definitions while teaching a model \citep{amershi2014power, gebreegziabher2025coadaptive, wang2021mutualtom}.

Taken together, these studies address coordination through models of human adaptation, shared mental models, or arbitration between human and machine policies, with smoother teamwork on a given task as the objective. Our framework differs in what is inferred and in how the interaction rule is obtained: the shared sign system itself is the latent variable inferred by the dyad, and the acceptance rule is derived from the generative model rather than designed. Neither agent's proposals are designated as authoritative, which distinguishes co-creative learning from both teacher-learner and guide-follower formulations. What emerges is therefore not only improved coordination but a symbol system, situating our framework within symbol emergence rather than task-level adaptation.

\subsection{Group-level Active Inference and Human-AI Symbiosis}
The extension of the Free Energy Principle (FEP) and active inference to multi-agent systems posits that a group can be modeled as a single agent minimizing its collective free energy \citep{taniguchi2025collective,balzan2023distributed}. A key feature of the CPC framework is its formalization of dual, interacting viewpoints: individual-level free energy minimization (akin to neural plasticity) and group-level minimization. As shown by Taniguchi et al.~\cite{taniguchi2025collective}, these are mediated by a collective regularization term in the energy function, which corresponds to the process of semiotic plasticity. Group-level active inference has been studied in research on cognitive niche construction \citep{Constant2018variational} and the emergence of cultural conventions \citep{Veissiere2020thinking}, where shared representations are built through the joint inference of interacting agents. Our concept of co-creative learning provides a concrete empirical instance of this process in the context of human-AI coexistence, demonstrating how a human-AI dyad can form a symbiotic cognitive system. This is crucial for designing future symbiotic societies by providing a mechanism for heterogeneous agents to establish the common ground needed for bidirectional alignment \citep{shen2024towards}. Our work addresses the challenge of integrating distributed information, a process also termed federated inference \citep{Friston2024federated}, by showing that an MHNG-based interaction provides a viable pathway towards such human-AI symbiosis.

\section{Preliminaries}
\label{sec:background}
\subsection{MHNG}
We consider a scenario based on the JA-NG \citep{okumura2023metropolishastings}, where two agents (indexed by $m \in \{A, B\}$) simultaneously attend to the same object $n$ (out of $N$ objects). Each agent possesses internal parameters $\Theta^m$ (representing knowledge about sign-concept mappings) and $\Phi^m$ (representing concept-observation mappings) and maintains an internal perceptual state (e.g., category assignment) $c_n^m$ for the object. The MHNG \citep{hagiwara2019symbol, taniguchi2023emergent} is a specific algorithm played within this JA-NG context. In each interaction concerning object $n$, one agent acts as Speaker (Sp) and the other as Listener (Li).
    \begin{enumerate}
        \item The Speaker samples a sign (name) $s_n^*$ based on its internal state and knowledge: $s_n^* \sim P(s_n|\Theta^{Sp}, c_n^{Sp})$.
        \item The Listener calculates the Metropolis-Hastings acceptance probability based on its own internal state $c_n^{Li}$, its currently associated sign $s_n^{Li}$, and the proposed sign $s_n^*$: 
        \begin{align}
r_n^{MH} = \min\left(1,\; \frac{P(c_n^{Li} \mid \theta^{Li}, s_n^*)}{P(c_n^{Li} \mid \theta^{Li}, s_n^{Li})} \right) \label{eq:mh_acceptance}.             
        \end{align}
        \item The Listener accepts the proposed sign $s_n^*$ (i.e., updates $s_n^{Li} \leftarrow s_n^*$) with probability $r_n^{MH}$.
        \item Agents update their internal parameters ($\Theta^m, \Phi^m, c_n^m$) via Bayesian inference (e.g., Gibbs sampling) conditioned on the interaction outcome. Roles are typically alternated.
    \end{enumerate}

    The MHNG process functions as a distributed implementation of an MCMC algorithm (specifically, Metropolis-Hastings sampling) targeting the posterior distribution over the shared signs $s_n$ conditioned on the distributed observations $x_n^A$ and $x_n^B$ of both agents \citep{hagiwara2019symbol, taniguchi2023emergent}. That is, the interaction allows the system to approximate samples from $P(\{s_n\}_{n=1}^N | \{x_n^A, x_n^B\}_{n=1}^N)$ without requiring agents to directly share their observations or internal parameters. This mechanism thereby enables the integration of distributed information. Theoretically, if human acceptance behavior in a JA-NG sufficiently aligns with the MH acceptance probability ($r_n^{MH}$), then the human-robot interaction can be viewed as performing decentralized Bayesian inference, leading to the emergence of shared signs ($s_n$) and integrated category representations ($c_n$) that reflect the integrated perceptual information, realizing co-creative learning. Okumura et al.~\cite{okumura2023metropolishastings} found empirical evidence supporting this alignment in human-human dyads.

\subsection{Interpersonal Gaussian Mixture (Inter-GM) Model} 
\begin{figure}[tb!p]
    \centering
  \includegraphics[width=\linewidth]{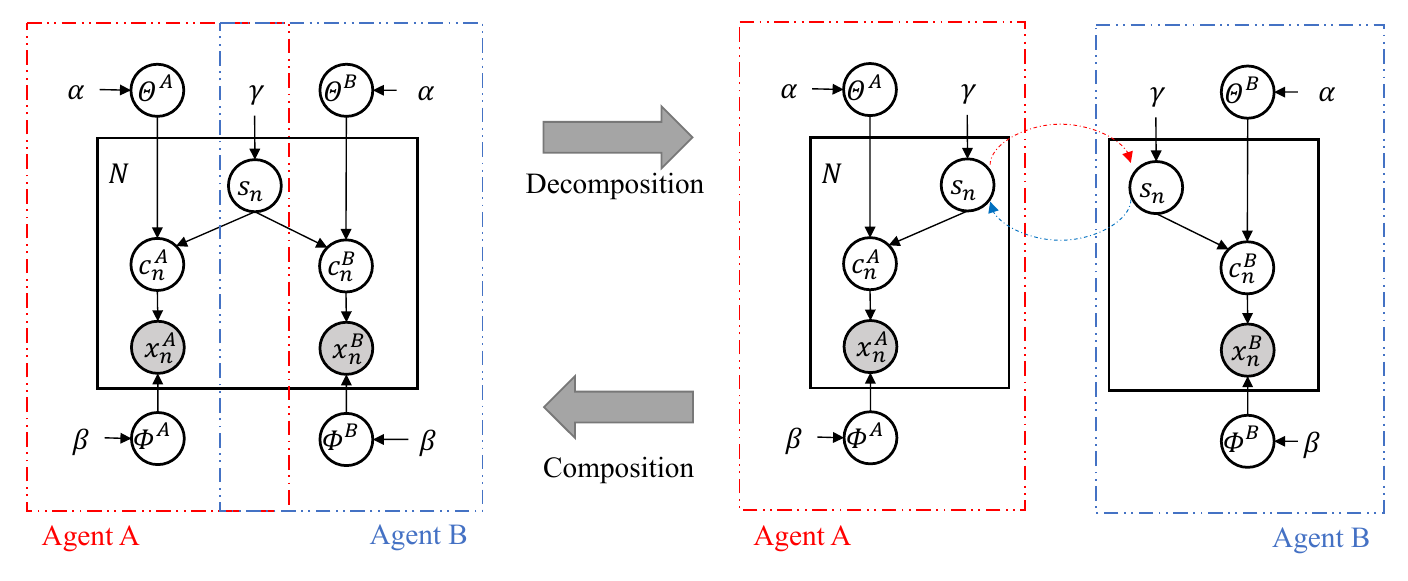}
  \caption{A typical probabilistic graphical model (left) assumed in CPC and its decomposition for decentralized inference through communication (right)~\citep{taniguchi2023emergent}.}
  \label{fig:inter_gm}
\end{figure}

For analysis and simulation, we employ the Inter-GM model, used in \cite{okumura2023metropolishastings}, which is a specific instance of the general Inter-PGM framework \citep{taniguchi2023emergent}. In Inter-GM, the observation $x_n^m$ for agent $m$ regarding object $n$ is assumed to be a continuous vector (here, features derived from color/shape). The internal perceptual state $c_n^m$ is a discrete latent variable representing the category assignment ($k \in \{1, ..., K\}$). The shared sign $s_n$ is also a discrete latent variable ($l \in \{1, ..., L\}$). The generative process assumes that $x_n^m$ is generated from a Gaussian distribution specific to its assigned category $c_n^m$, parameterized by $\phi_k^m = \{\mu_k^m, \Lambda_k^m\}$. The category $c_n^m$ is generated from a categorical distribution specific to the shared sign $s_n$, parameterized by $\theta_l^m = \{\theta_{l,k}^m\}_{k=1}^K$. The sign $s_n$ is drawn from a prior $\pi$. Figure~\ref{fig:inter_gm} illustrates the probabilistic graphical model for the Inter-GM, showing both the conceptual integration of the two agents' models via the shared sign $s_n$ (Left) and its decomposition for decentralized inference (Right), which forms the basis for algorithms like MHNG. 

    In our analysis, the internal parameters $\theta^m$, $\phi^m$, and category assignments $c_n^m$ for each agent (human participant analysis or computer agent simulation) are inferred using Gibbs sampling conditioned on the observed data $x_n^m$ and the sequence of accepted signs $s_n$ obtained through the JA-NG interactions. The MH acceptance probability $r_n^{MH}$ (Eq.~\ref{eq:mh_acceptance}) is calculated using these inferred parameters.

\section{Co-Creative Learning}

\subsection{Conceptual Formulation in Contrast to Traditional Paradigms}

To clarify our framework, we first define the basic setup. Let two agents, $\mathcal{A}=\{\text{AI, Human}\}$, each observe partial information about a shared world. For each data instance $n=1,...,N,$ they receive respective observations $x_n^m$ ($m\in\mathcal{A}$).
With this setup, the unique positioning of co-creative learning can be clarified by formalizing the flow of information in contrast to traditional machine learning paradigms.

In a typical \textbf{supervised learning} scenario, a human agent observes data, such as an image $x_n^\text{Human}$, and provides a corresponding ground-truth label $s_n$. This process effectively defines a posterior distribution $p(s_n | x_n^\text{Human})$\footnote{Note that the posterior distribution reflects the symbol system that human society, i.e., a {\emph symbol emergence system},  collectively forms~\citep{taniguchi2024collective}).}, which is treated as the target for the AI to learn. The AI, with its own observation $x_n^\text{AI}$, is then trained to have its approximate posterior $q(s_n | x_n^\text{AI})$ mimic the human's, such that $q(s_n | x_n^\text{AI}) \approx p(s_n | x_n^\text{Human})$. From the perspective of symbol emergence, the AI's understanding of the world is fundamentally bounded by the human's perspective, $p(s_n | x_n^\text{Human})$.

In a typical \textbf{unsupervised learning}, an AI agent learns in isolation, attempting to infer a posterior distribution $p(s_n | x_n^\text{AI})$ solely from its own observations. In this case, its understanding is bounded by the limits of its own perceptual data, $x_n^\text{AI}$. Indeed, much research in representation learning, symbol emergence, and cognitive developmental robotics has taken this approach, where an agent performs object category formation, concept learning, or world model learning from raw image or multimodal information, often grounded in the principles of predictive coding~\citep{kingma2014auto,suzuki2022survey,schmidhuber1990making,ha2018world,friston2021world,taniguchi2016symbol,taniguchi2019symbol,taniguchi2023world}.

\textbf{Co-creative learning} presents a third approach. Instead of treating one agent's perspective as the ground truth, the learning target for the dyad is the true posterior conditioned on the \textbf{joint observations} of both agents: $p(s_n | x_n^\text{Human}, x_n^\text{AI})$. The goal is for the human-AI system, through interaction, to collectively approximate this richer, more informed posterior. Because this posterior is conditioned on both agents' observations, the shared representation that the dyad converges on reflects information that neither the human nor the AI could have recovered from its own partial observation alone. We use this operational sense throughout: the claim concerns the completion of complementary information under partial observability, not the emergence of a representation that is qualitatively richer, such as one exhibiting higher-order abstraction or compositional structure. This approach is grounded in the principles of CPC, which aims to achieve symbol emergence through decentralized Bayesian inference within a heterogeneous system of humans and AI~\citep{taniguchi2024collective}.

\subsection{Generative Model and Formal Definition}
\label{sec:formal_def}

The observations $x_n^m$ ($m\in\mathcal{A}$) are assumed to be generated from a common latent symbol $s_n$ that represents the shared, external representation:
\begin{equation}
s_n\sim p(s|\gamma), \quad x_n^{m}\sim p(x|s_n,\Theta^m) \quad (m\in\mathcal{A}).
\label{eq:gen_model}
\end{equation}
Here, $\Theta^{m}=\{\phi^{m},\theta^{m}\}$ denote the parameters for each agent, and $\gamma$ is a hyperparameter. Notably, the observations are often structured via agent-specific internal representations, where a latent internal concept $c_n^{m}$ mediates the perception process: $c_n^{m}\sim p(c | s_n,\theta^{m})$ and $x_n^m\sim p(x|c_n^{m},\phi^{m})$.

Based on this generative process, we can now formally define co-creative learning:

\begin{mdframed}[backgroundcolor=gray!4!white,
                 linewidth=0.8pt,
                 roundcorner=4pt]
\textbf{Definition 1 (Co-Creative Learning)} Let $X_n=(x_{n}^{m})_{m\in\mathcal{A}}$ be the partial observations generated via the shared latent symbol $s_{n}$ as defined above. Co-creative learning is the interactive, decentralized Bayesian inference process in which the agents, through message-based updates (e.g., the Metropolis-Hastings naming game), jointly construct local belief distributions $q_{t}^{m}(s_{n})$ such that the collective free energy,
\begin{align*}
    \mathcal{F}_{t} &= -\log p(X^\text{AI},X^\text{Human}) + \text{KL}(q_{t}(s)||p(s|X^\text{AI},X^\text{Human})), \\
    \text{with} \quad q_{t}(s) &\propto \prod_{m\in\mathcal{A}}q_{t}^{m}(s),
\end{align*}
satisfies
\[ \mathbb{E}[\mathcal{F}_{t+1}]\le\mathbb{E}[\mathcal{F}_{t}]. \]
That is, the expected collective predictive-coding free energy decreases over interaction steps. Equivalently, the Markov chain formed by the interactions converges (in distribution) to the true posterior $p(s|X^\text{AI}, X^\text{Human}, \Theta)$, where $\Theta=(\Theta^{m})_{m\in\mathcal{A}}$, so that Monte Carlo estimates of the evidence $p(X^\text{AI},X^\text{Human})$ improve without requiring either agent to disclose private observations or gradients.
\end{mdframed}

Three terms should be distinguished here, and we use them consistently in what follows: \emph{observation} refers to the private sensory input $x_n^m$, \emph{belief} to an agent's posterior over the latent variables $c_n^m$ and $s_n$, and \emph{knowledge} to the model parameters $\Theta^m$. The clause ``without requiring either agent to disclose private observations or gradients'' is an assumption of infeasibility rather than a prohibition: nothing in the protocol forbids disclosure, but full and faithful disclosure is generally unavailable because of cognitive closure between agents. An AI cannot directly access a human's internal perceptual and cognitive states. Conversely, a human has only limited access to an AI's internal states, because parameters and high-dimensional latent representations are not directly interpretable even when exposed. A framework for human-AI collaboration should therefore not depend on full disclosure. It is precisely because observations cannot be shared directly that the shared sign $s_n$ must carry the burden of integration, which is why symbol emergence is required at all in this setting.

The non-increasing property of the collective free energy, $\mathbb{E}[\mathcal{F}_{t+1}]\le\mathbb{E}[\mathcal{F}_{t}]$, is a direct consequence of the mechanics underlying the MHNG. The protocol effectively turns the dyadic interaction into a distributed Metropolis-Hastings sampler, a type of Markov chain Monte Carlo (MCMC) algorithm specifically designed to converge to a target distribution. The key to this convergence is that the transition kernel of the chain satisfies the detailed balance condition with respect to the true posterior $p(s|X^{\text{AI}}, X^{\text{Human}})$. The satisfaction of detailed balance mathematically guarantees that the KL-divergence between the agents' collective belief $q_t(s)$ and the true posterior is monotonically non-increasing in expectation. Since the collective free energy $\mathcal{F}_t$ is defined by this KL-divergence (up to an additive constant), it inherits this essential property. This mechanism is a well-established result in information theory, often referred to as an instance of the Data Processing Inequality~\citep{Cover2006}.

The collective free energy $\mathcal{F}_{t}$ presented here is a specialized formulation for a two-agent human-AI system, derived from the CPC framework proposed by Taniguchi et al. \cite{taniguchi2025collective}. The CPC framework, in turn, generalizes the free energy principle for individual agents \citep{friston2010free} to the collective, multi-agent level to account for the symbol emergence.

It is worth stating explicitly what is and is not given in advance in this formulation, since this determines the sense in which the process is creative. The symbol $s_n$ is a latent variable: which signs come to stand for which regions of the observation space, and the sign-to-concept mapping $\theta^m$ that gives those signs their meaning for each agent, are not specified beforehand. They are outcomes of the interaction. In the experiment reported below, the ground truth categories serve only as an external yardstick for our evaluation; they are available to neither agent and they do not determine the signs that emerge. What the dyad produces is therefore a shared sign system that neither participant possessed in advance and that neither could have derived from its own observations alone. We use the term co-creative in this specific sense: the joint construction of a shared symbol system through decentralized inference, which is the sense of symbol emergence that our framework addresses \citep{taniguchi2024collective, taniguchi2026generative}.\footnote{The present study is deliberately a minimal instantiation. To obtain a controlled test of whether human-AI dyads can engage in this process at all, we hold the concept space, the modality, and the artifact type fixed. Richer forms of co-creative process arising from the same underlying principle are being investigated in adjacent studies: joint composition across musical genres via a Metropolis-Hastings music game \citep{sakurai2026mhmug}, knowledge fusion of vision-language models via a Metropolis-Hastings captioning game \citep{matsui2025mhcg}, and the co-construction of the social reality of emotion \citep{nomura2026socialemotion}. Computational models of artificial-agent interaction have already exhibited richer outcomes, including compositional generalization and emergent musical creativity \citep{lehoang2026compositional,sakurai2026mhmug}. This framing is consistent with system-theoretic accounts of creativity, in which a creative act is a co-construction between a generative proposal and its reception \citep{csikszentmihalyi1996creativity}, a structure that recurs in music, in literature, and in scientific inquiry \citep{taniguchi2025cpcms, taniguchi2025system0123}. Co-creative learning as instantiated here is a minimal model of that structure.}

\subsection{Implications for Human-AI Symbiosis}
This formulation provides a theoretical basis for designing AI systems that learn collaboratively \emph{with} humans, rather than learning unilaterally \emph{from} them in a supervised fashion. The process is one of joint inference over shared external representations (symbols), where knowledge is co-constructed instead of simply transferred. \deleted{By minimizing the collective free energy through message-passing dynamics, co-creative learning realizes a form of bidirectional, symbiotic AI alignment grounded in mutual perception \citep{shen2024towards}.}\added{By minimizing the collective free energy through message-passing dynamics, co-creative learning is hypothesized to provide a basis for a form of bidirectional, symbiotic AI alignment grounded in mutual perception \citep{shen2024towards}.} The emergence of a shared symbol system is the core mechanism for establishing the "common ground" upon which this symbiosis between heterogeneous agents is built.

In practice, the framework's decentralized nature is a key advantage, allowing for dynamic, situated interactions that update both parties' internal models while respecting the cognitive boundaries of each agent (i.e., without requiring access to private internal states or gradients). In the following sections, we demonstrate this process empirically using the JA-NG, where a human participant and an AI agent play MHNG implicitly and explicitly, respectively, as a concrete instantiation.

\section{Material and Methods}
\label{sec:experiment}

\textbf{Objective:}
    \label{subsec:exp_objective:} 
    The primary objective of this experiment was to investigate whether human-robot pairs could engage in co-creative learning through a JA-NG under partial observation conditions. We operationally define co-creative learning as the ability of the dyad to integrate their disparate perceptual information to achieve better-than-individual categorization accuracy (measured by ARI relative to ground truth) and converging on shared signs mapping to these integrated categories. A secondary objective was to evaluate the effectiveness of the MHNG acceptance mechanism in facilitating this process, by comparing it against baseline acceptance strategies (Always Accept, Always Reject).

\textbf{Participants and Ethics:} 
    We recruited 90 participants via the crowdsourcing platform CrowdWorks. Participants whose interaction logs indicated inactivity or failure to follow instructions (e.g., never changing classifications or sign assignments throughout the experiment) were excluded, resulting in a final analysis set of $N=69$ participants (MH group: $26$, AA group: $20$, AR group: $23$). Participants received $2000$ JPY (approx. \$13 USD in late 2023/early 2024) for their time (approx. $90-120$ minutes).

 The study protocol was approved by the Research Ethics Committee for Non-Medical Life Science Research at Ritsumeikan University (Approval No. BKC-LSMH-2023-095). All participants provided informed consent online via the experimental web application before starting the experiment. Participation was voluntary, and participants could withdraw at any time without penalty. Further details are provided in Appendix~\ref{sec:appendix_ethics}.

\textbf{Task and Stimuli:}
    The stimuli consisted of images depicting circles with a radial notch ($N=10$ distinct stimuli per round). These were generated based on 5-dimensional vectors $\mathbf{x}_n = (L^*, U^*, V^*, \text{Angle}, \text{Size})^\top$ sampled from one of $K=3$ ground truth Gaussian distributions $\mathcal{N}(\mathbf{x} | \boldsymbol{\mu}_k, \boldsymbol{\Sigma})$. The first three dimensions represent color in the perceptually uniform CIE-L*U*V* space \citep{steels2005coordinating}. The parameters ($\boldsymbol{\mu}_k, \boldsymbol{\Sigma}$) were chosen such that the categories overlap significantly in the partial observation spaces.

\begin{figure}[tb!p]
    \centering
    \subfloat[Dataset and partial observation setup.]{\includegraphics[width=0.5\linewidth]{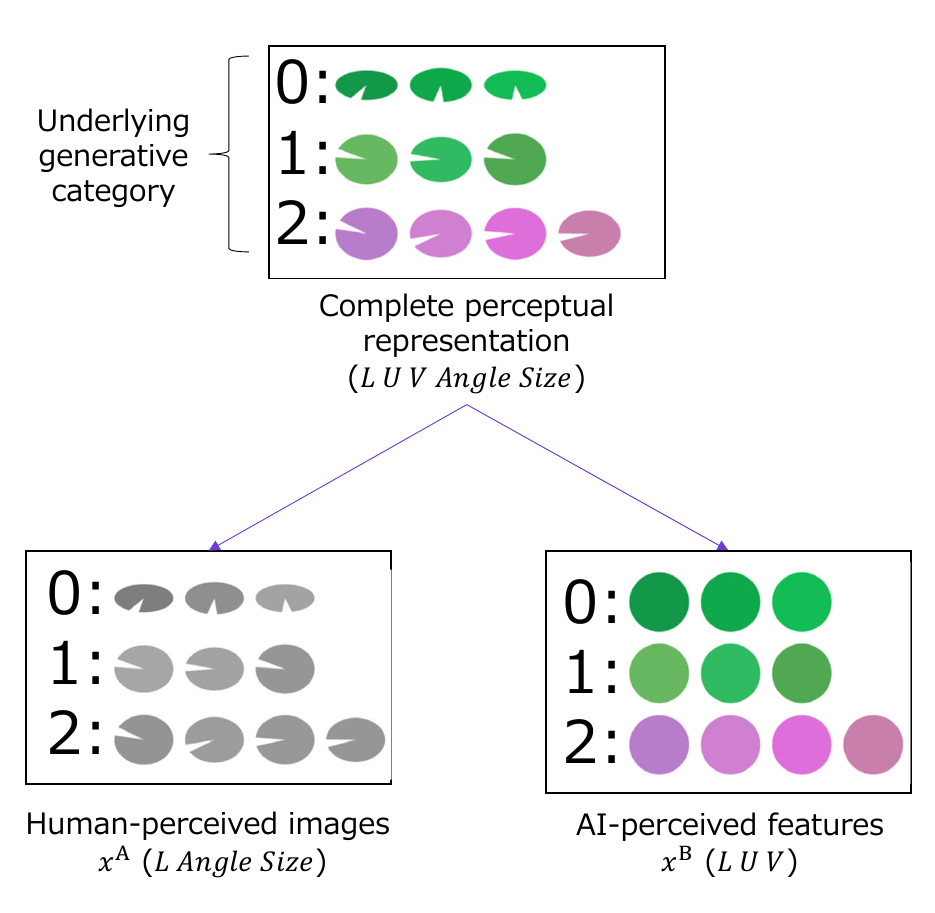}\label{fig:partial_observation}}
    \vspace{1em} 
    \subfloat[Experimental procedure flow (JA-NG).]{\includegraphics[width=0.8\linewidth]{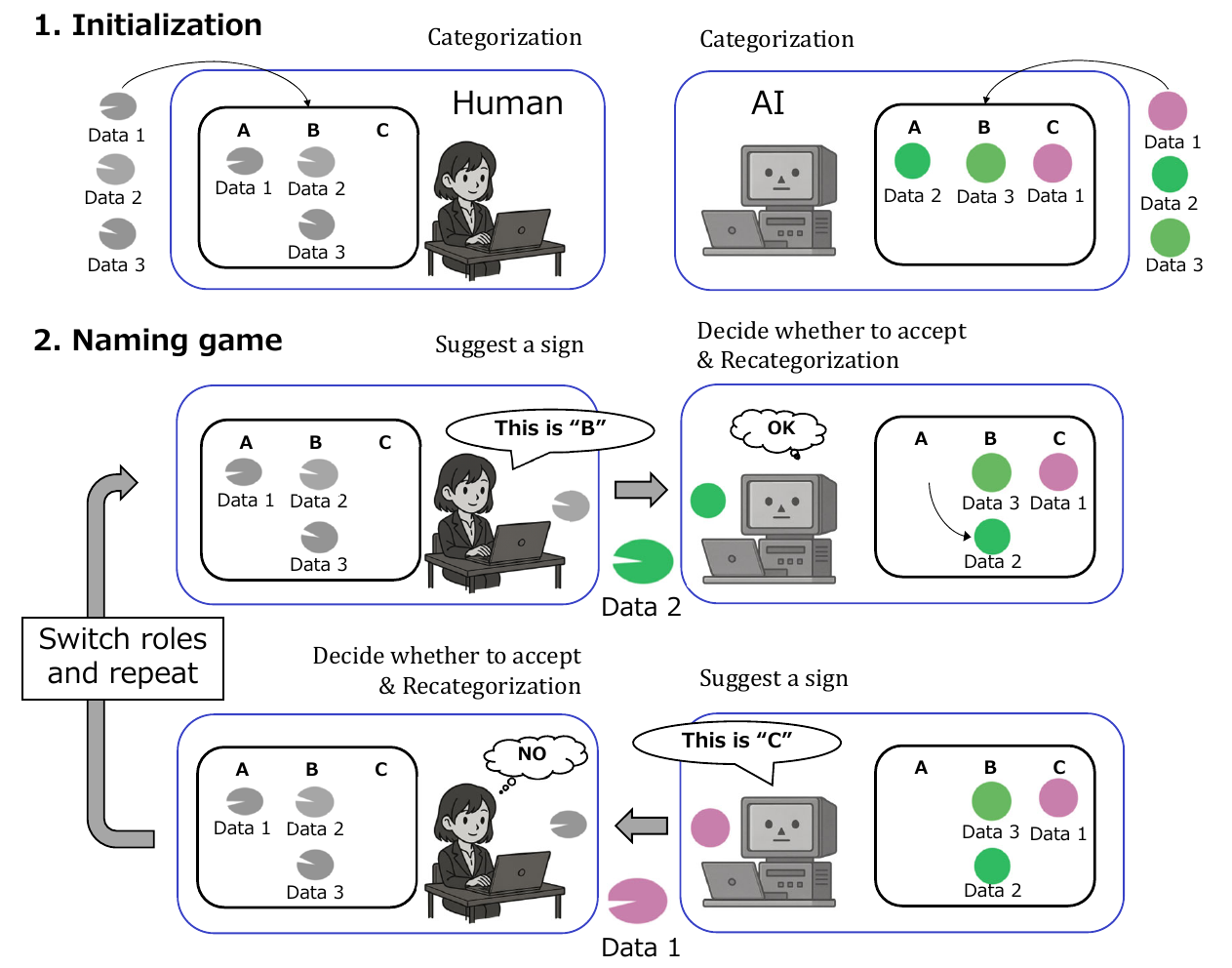}\label{fig:experiment_flow}}
    \caption{Experimental setup: (a) Stimuli under partial observation and (b) JA-NG procedure flow.}
    \label{fig:exp_setup_combined}
\end{figure}

    Humans and the computer agent received different partial observations. Human participants viewed a grayscale rendering based on ($L^*$, Angle, Size) dimensions. The computer agent received the numerical values for ($L^*, U^*, V^*$) dimensions. Figure~\ref{fig:partial_observation} illustrates this setup. This partial observability makes accurate categorization impossible based solely on individual information, necessitating information integration through communication.
    
    The dyad's task was to collaboratively categorize the 10 stimuli into one of up to three categories (using abstract labels A, B, C) by playing the JA-NG. Participants were instructed to cooperate with their computer partner to improve classification accuracy.

    \textbf{Procedure and Conditions:}
The experiment was conducted online using a custom web application developed with the oTree framework~\cite{chen2016otree}.
 After consent and instructions, participants performed an initial categorization of the 10 stimuli into $\le 3$ categories (A, B, C). The computer agent also initialized its categorization using its Inter-GM model based on its partial view. Participants then engaged in 20 rounds of the JA-NG. Each round comprised 10 naming interactions (one for each stimulus, presented in random order), totaling 200 interactions per participant.
    In each naming interaction (illustrated in Fig~\ref{fig:experiment_flow}), the speaker (human or computer, implicitly alternating) proposed a name (A, B, or C) for the current stimulus. The listener decided to `accept` or `reject` via button press (human) or according to its condition's rule (computer). Listeners could update their categorization at any time, prompted after an accept/reject decision.
    Participants were randomly assigned to one of three groups, differing in the computer agent's acceptance strategy for names proposed by the human:
        \begin{itemize}
            \item \textbf{Group 1 (MH):} AI accepts human's proposal with probability $r_n^{MH}$ (Eq.~\ref{eq:mh_acceptance}).
            \item \textbf{Group 2 (Always Accept; AA):} AI always accepts human's proposal.
            \item \textbf{Group 3 (Always Reject; AR):} AI always rejects human's proposal.
        \end{itemize}

These three conditions were designed to correspond to the three learning paradigms discussed in Section 2.1. The {\bf AA (Always Accept)} condition serves as a proxy for {\bf supervised learning}; in this setup, the AI passively accepts all labels proposed by the human, effectively being unilaterally 'taught' based on the human's perspective ($p(s_n|x_n^\text{Human})$). \deleted{Conversely, the {\bf AR (Always Reject)} condition mirrors {\bf unsupervised learning}, where the AI forms categories based solely on its own perceptual data ($p(s_n|x_n^\text{AI})$).}\added{Conversely, the {\bf AR (Always Reject)} condition represents {\bf unsupervised learning}, i.e., unsupervised perceptual category formation: the AI infers its internal latent variable $s_n$ so as to increase the predictability of $p(x_n^\text{AI})$, without receiving top-down class labels or incorporating information from its human partner.}\footnote{\added{Here, unsupervised learning refers to modeling an input distribution by inferring latent variables from unlabeled perceptual information, e.g., forming categories, as commonly studied in symbol emergence in robotics \citep{taniguchi2024collective,taniguchi2016symbol}. This characterization is from the viewpoint of perceptual categorization; how the phenomenon as a whole is framed depends on the theoretical framework through which it is viewed.}} Finally, the central {\bf MH} condition is designed to instantiate {\bf co-creative learning}, where the human and AI mutually influence each other's representations through a process of decentralized Bayesian inference that realizes symbol emergence via CPC and aims to approximate the joint posterior ($p(s_n|x_n^\text{Human}, x_n^\text{AI})$).

We emphasize that AA and AR are limiting cases rather than models of realistic collaboration. In particular, an agent that categorically rejects every proposal is not a plausible interaction partner; AR is included as a \deleted{lower-bound control that isolates the unsupervised extreme}\added{lower-bound control for unsupervised learning, i.e., unsupervised perceptual category formation}, in which no information from the partner enters the AI's inference. We adopted this three-condition design because our central claim concerns the theoretical status of co-creative learning as a paradigm distinct from supervised and unsupervised learning, and the paradigm-defining limits are what test that claim most directly. We also note a constraint specific to studies of this kind: because co-creative learning is defined by the AI's decision policy interacting with a human in real time, a fixed dataset cannot be reused across policies, and each additional condition requires an additional independent group of participants. This bounds the number of conditions that can be compared within a single study. Baselines with stochastic or partially reliable acceptance would be informative, and we return to them in Section~\ref{sec:discussion} as the natural next step.

    The computer always proposed names based on its internal state. Humans always made their own acceptance decisions.

    \textbf{Evaluation Metrics:} 
    We measured categorization accuracy using the Adjusted Rand Index (ARI) \citep{hubert1985comparing}, comparing the category assignments ($c_n^*$) of both the human and the computer agent against the ground truth ($K=3$) category labels at each communication step. ARI ranges from -0.5 (worse than random) to 1 (perfect agreement), with 0 indicating chance performance.
    To assess convergence towards a shared and informative sign system, we measured the agreement between the empirical sign distribution (over $s_n \in \{A, B, C\}$ for each object $n$, estimated from the last 5 rounds, i.e., 50 interactions) for each agent and a target distribution. The target distribution was approximated by the sign posterior derived from Gibbs sampling on the full Inter-GM model given both human and computer observations $\{x_n^A, x_n^B\}_{n=1}^N$ (averaged over 100 independently generated Gibbs-sampling histograms). Agreement was calculated as the normalized sum of minimum histogram counts for each sign assignment across all objects, after optimally matching sign labels between the empirical and target distributions using the Hungarian algorithm on a cost matrix derived from pairwise minimum overlaps \citep{inukai2023recursive}. This agreement score ranges from 0 (no overlap) to 1 (identical distributions).

    \textbf{Modeling Human Acceptance Behavior:} 
    To quantify the influence of the theoretical MH acceptance probability ($r_n^{MH}$) on human decisions, we modeled the human probability of accepting ($z_n=1$) a computer's name proposal using a constrained linear Bernoulli model: $P(z_n=1 | r_n^{MH}) = a r_n^{MH} + b$, s.t. $0 \le b$ and $a+b \le 1$. For each instance where the human acted as listener, $r_n^{MH}$ was calculated using the human's inferred Inter-GM parameters ($\Theta^A$) and category assignment ($c_n^A$) via Gibbs sampling. The parameters $a$ (sensitivity to $r_n^{MH}$) and $b$ (baseline acceptance) were estimated via MLE across all relevant interactions for all participants using gradient descent.

\textbf{Statistical Testing:} For each outcome we first tested for any difference among the three groups using the Kruskal-Wallis test, followed by Dunn's post hoc test. The three pairwise condition contrasts following one outcome-specific omnibus test were treated as a single family, and Holm correction was applied to those three comparisons. We chose a nonparametric procedure because the per-participant ARI and agreement scores are bounded and not well approximated as normal. The significance level was set at $\alpha=0.05$. We report all five outcomes and interpret evidence across outcomes cautiously, since the outcome families were not preregistered and no comparison was designated as confirmatory. In a sensitivity analysis applying Holm correction across all fifteen outcome-wise pairwise comparisons as a single family, the adjusted $p$-value for the computer sign ARI comparison between MH and AR was $0.075$, whereas the computer agreement comparison between MH and AR remained significant ($p = 1.3 \times 10^{-9}$).

\section{Results}
\label{sec:results}
\begin{figure}[tb!p]
      \centering
      \includegraphics[width=\linewidth]{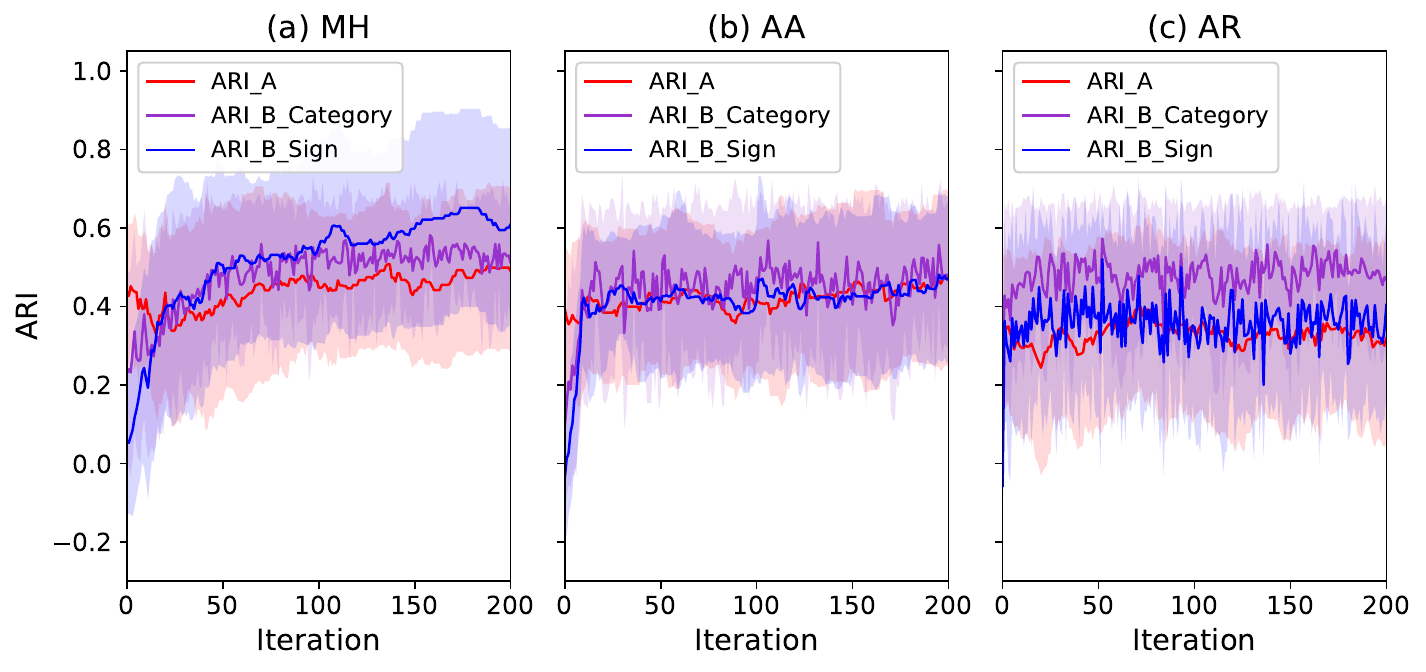}
        \caption{Mean ARI ($\pm$ SD) over 200 steps for Human and Computer agents under MH, Always Accept (AA), and Always Reject (AR) conditions. The shaded bands show the sample standard deviation across participants ($N = 26$, $20$, and $23$ for MH, AA, and AR, respectively), so that the dispersion of individual outcomes is displayed directly.}
      \label{fig:ari_results}
    \end{figure}

   \textbf{Clustering Accuracy (ARI):} 
We present the results based on the filtered dataset (N=69 participants: 26 in MH, 20 in AA, 23 in AR).
    Figure~\ref{fig:ari_results} shows the evolution of the mean ARI over the 200 communication steps for both humans and computer agents in the three conditions. Both the MH (Group 1) and AA (Group 2) conditions show a general trend of increasing ARI for both humans and computers, indicating successful information integration and learning from the interaction. In contrast, the AR (Group 3) condition shows little to no improvement from the initial categorization accuracy, especially for the computer agent.

    Table~\ref{tab:merged_results} summarizes the initial and final mean ARI scores.

Table~\ref{tab:posthoc} reports the omnibus tests and the outcome-specific Holm-corrected pairwise comparisons for all measures. For the sign-based ARI of the computer agent, the omnibus test indicated a difference among the three groups ($H = 7.89$, $p = 0.019$). The MH condition attained the highest final value ($0.609 \pm 0.251$) and exceeded the AR condition ($0.404 \pm 0.228$) after outcome-specific Holm correction (Dunn's test, $p = 0.017$). This difference did not survive Holm correction when all fifteen pairwise comparisons were treated as one family ($p = 0.075$); we therefore regard it as suggestive rather than decisive. The difference from the AA condition ($0.469 \pm 0.212$) did not survive outcome-specific Holm correction ($p = 0.173$). For the human participants' category ARI, the omnibus test was at the threshold of significance ($H = 5.87$, $p = 0.053$); the MH condition reached $0.490 \pm 0.208$ against $0.473 \pm 0.224$ for AA and $0.319 \pm 0.267$ for AR, and no pairwise difference survived outcome-specific Holm correction (MH vs. AR, $p = 0.077$). For the computer agent's category ARI, the omnibus test showed no difference ($H = 0.74$, $p = 0.690$). Thus, the sign-based ARI provides suggestive evidence that interaction reshapes the shared sign system rather than either agent's private categories, while the most robust evidence for integration is the agreement with the joint-observation posterior reported below.

\begin{table}[bt]
 \tbl{Mean ARI and Sign Agreement Scores ($\pm$ SD) across groups and phases (N=69).}
  {\begin{tabular}{@{}l l c c c c@{}}
    \toprule
    Group (N agents) & Phase & Human category ARI & Computer sign ARI & Human Agree. & Computer Agree. \\
    \midrule
    G1 (MH, n=26) & Initial & $0.438 \pm 0.165$ & $0.055 \pm 0.180$ & \multirow{2}{*}{$0.728 \pm 0.067$} & \multirow{2}{*}{$\mathbf{0.765 \pm 0.070}$} \\
                  & Final   & $\mathbf{0.490 \pm 0.208}$ & $\mathbf{0.609 \pm 0.251}$ & & \\
    \addlinespace
    G2 (AA, n=20) & Initial & $0.394 \pm 0.145$ & $-0.038 \pm 0.114$ & \multirow{2}{*}{$0.722 \pm 0.075$} & \multirow{2}{*}{$0.717 \pm 0.075$} \\
                  & Final   & $\mathbf{0.473 \pm 0.224}$ & $0.469 \pm 0.212$ & & \\
    \addlinespace
    G3 (AR, n=23) & Initial & $0.344 \pm 0.200$ & $-0.057 \pm 0.109$ & \multirow{2}{*}{$0.650 \pm 0.121$} & \multirow{2}{*}{$0.468 \pm 0.055$} \\
                  & Final   & $0.319 \pm 0.267$ & $0.404 \pm 0.228$ & & \\
    \bottomrule
  \end{tabular}}
 \label{tab:merged_results}
\end{table}

\begin{table}[bt]
  \centering
  \tbl{Omnibus tests and outcome-specific Holm-corrected post hoc comparisons on the final scores ($N=69$: MH $26$, AA $20$, AR $23$). Omnibus: Kruskal-Wallis. Post hoc: Dunn's test with Holm correction across the three contrasts for each outcome. Boldface marks comparisons significant at $\alpha=0.05$ under this outcome-specific correction. For the two marked MH--AR contrasts, a global sensitivity analysis treating all fifteen pairwise comparisons as one family yielded $p=0.075$ for computer sign ARI$^\dagger$ and $p=1.3\times10^{-9}$ for computer agreement$^\ddagger$.}
  {\small
  \begin{tabular}{@{}l c c c c@{}}
    \toprule
    Measure & Kruskal-Wallis & MH vs. AA & MH vs. AR & AA vs. AR \\
    \midrule
    Human category ARI            & $H=5.87$, $p=0.053$ & $0.871$ & $0.077$ & $0.107$ \\
    Computer category ARI         & $H=0.74$, $p=0.690$ & $1.000$ & $1.000$ & $1.000$ \\
    Computer sign ARI    & $H=7.89$, $p=0.019$ & $0.173$ & $\mathbf{0.017}^{\dagger}$ & $0.360$ \\
    Human agreement      & $H=5.93$, $p=0.052$ & $0.848$ & $0.073$ & $0.109$ \\
    Computer agreement   & $H=45.87$, $p=1.1\times10^{-10}$ & $0.227$ & $\mathbf{2.6\times10^{-10}}^{\ddagger}$ & $\mathbf{1.9\times10^{-6}}$ \\
    \bottomrule
  \end{tabular}}
  \label{tab:posthoc}
\end{table}

 \textbf{Sign Posterior Distribution Convergence}
We evaluated the convergence towards a shared and informative sign system by measuring the agreement between each agent's final sign distribution and the target posterior distribution derived from the combined model. 
Table~\ref{tab:merged_results} shows the mean agreement scores.

\textbf{Why the sign-based ARI rises under MH.} The contrast between the sign-level and category-level results identifies where integration takes place, and it follows from the mechanism. The sign $s_n$ held by the AI is neither a label transmitted by the human nor a label formed from the AI's own observation: its update is governed by the MH acceptance rule, which is precisely the decentralized sampler for the posterior conditioned on both agents' observations. The sign trajectory is therefore driven toward the partition supported by the union of the two partial observations, which lies closer to the ground truth partition than either agent's own observation supports. The measurements bear this out: the AI's sign-based ARI rises from $0.055 \pm 0.180$ to $0.609 \pm 0.251$ under MH, whereas its category-level ARI ends at $0.511 \pm 0.161$ and does not differ across conditions ($p = 0.690$). The private perceptual models are largely unchanged by the interaction; the shared sign system is where the integration is realized. This is the signature predicted by the collective predictive coding account, in which integration is carried by the external representation rather than by either agent's internal model. We note that this systematic rise coexists with substantial dispersion across participants: the sample standard deviation of the final sign-based ARI is $0.251$ under MH, against $0.212$ under AA and $0.228$ under AR. The outcome-specific comparison supports this pattern, but its significance is sensitive to the correction family, as reported above. Fig.~\ref{fig:ari_results} displays sample standard deviations rather than standard errors.

The two agents also occupy structurally different positions, and this asymmetry has a visible consequence. The AI evaluates the acceptance test using an explicit generative model and therefore realizes the sampler exactly, whereas the human approximates it, as the acceptance-behavior analysis below indicates. The component that most faithfully implements the decentralized inference is thus the AI's sign variable, which is consistent with the effect being strongest and statistically robust on the AI side ($p = 2.6 \times 10^{-10}$ against AR for sign agreement) while the corresponding human-side differences are weaker ($p = 0.073$).

For the computer agent, the omnibus test showed a clear difference among the groups ($H = 45.87$, $p = 1.1 \times 10^{-10}$). Agreement was highest in the MH condition ($0.765 \pm 0.070$) and far exceeded the AR condition ($0.468 \pm 0.055$) after outcome-specific Holm correction (Dunn's test, $p = 2.6 \times 10^{-10}$); this difference remained significant under the global-family correction ($p = 1.3 \times 10^{-9}$). The AA condition ($0.717 \pm 0.075$) also exceeded AR ($p = 1.9 \times 10^{-6}$; global-family $p = 1.3 \times 10^{-5}$). The difference between MH and AA did not survive outcome-specific Holm correction ($p = 0.227$). For the human participants, the omnibus test was at the threshold ($H = 5.93$, $p = 0.052$), with MH at $0.728 \pm 0.067$, AA at $0.722 \pm 0.075$, and AR at $0.650 \pm 0.121$, and no pairwise difference surviving outcome-specific Holm correction. The computer agreement result therefore provides the most robust evidence that MH guides the AI toward a sign system aligned with the joint-observation posterior more effectively than self-directed learning under AR. No statistically significant difference from unconditional acceptance under AA was detected.

\begin{figure}[tb!p]
  \centering
  \includegraphics[width=0.40\textwidth]{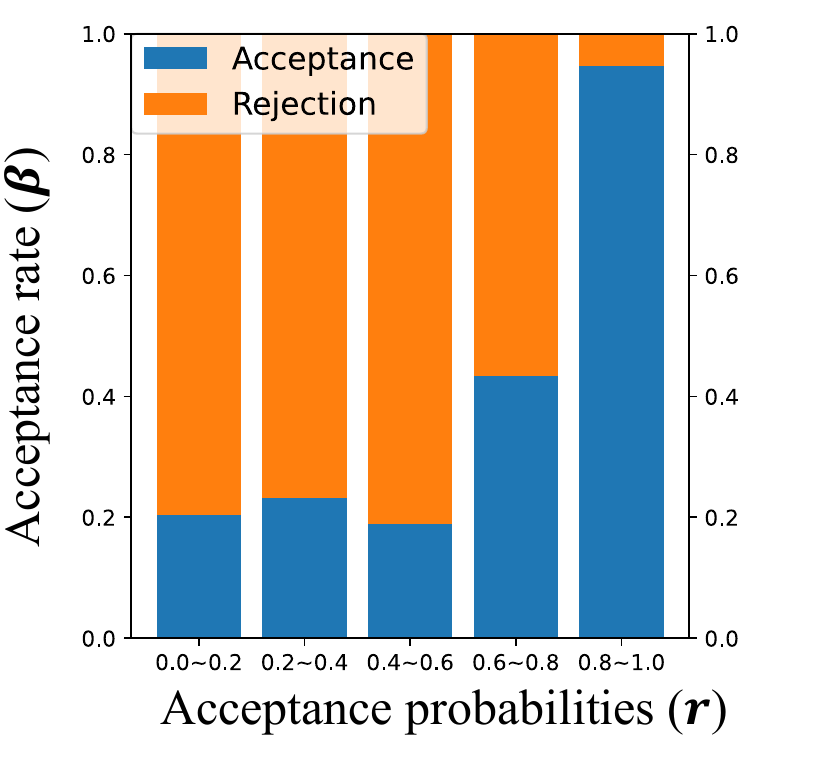}
  \caption{Human acceptance rate for binned $r_n^{MH}$ (N=69).}
  \label{fig:human_acceptance}
\end{figure}

    \textbf{Analysis of Human Acceptance Behavior}
    \label{subsec:results_human}
    Figure~\ref{fig:human_acceptance} plots the actual acceptance rate of human participants against the binned MH acceptance probability ($r_n^{MH}$) calculated based on their inferred internal state. A clear positive correlation is visible: humans were more likely to accept the computer's proposal when the MH probability was high, and less likely when it was low. The fitted linear model yielded parameters $\hat{a} = 0.645 \pm 0.300$ and $\hat{b} = 0.201 \pm 0.187$ (mean $\pm$ SD across participant).

Furthermore, the analysis of human acceptance decisions provides additional support for the underlying mechanism. We observed a strong positive relationship between the theoretical MH acceptance probability ($r_n^{MH}$) and the actual choices made by human participants, with the fitted parameter $\hat{a}$ being clearly positive. 

This observation is consistent with MHNG theory, and it extends the human-human finding of \cite{okumura2023metropolishastings} to a setting whose structure differs in a way that matters. In the human-human case, both parties are approximate Bayesian reasoners of the same kind, and a correspondence between acceptance behavior and the theoretical probability may be attributed to a symmetry between two similar cognitive systems. Removing that symmetry makes the present result informative in three respects. First, the partner here computes its acceptance probability exactly from an explicit generative model to which the human has no access, so the correspondence cannot be explained by the two parties running comparable machinery; that human acceptance nonetheless tracks the derived probability is not implied by the human-human case. Second, the dyad is consequently a hybrid of an exact and an approximate sampler, and our results show that such a hybrid still converges on a shared sign system aligned with the joint posterior. Third, the practical import differs: in the human-human case the finding characterizes human behavior, whereas here it establishes that an AI can be designed to participate in this process on principled grounds, since its acceptance rule is derived rather than tuned, and that humans engage with such an agent in the manner the theory predicts.

\section{Discussion}
\label{sec:discussion}

Our experiments show that human-robot pairs can successfully engage in co-creative learning via JA-NG under partial observation. Interaction improved categorization accuracy (ARI), especially in the MH and AA conditions (Table~\ref{tab:merged_results}, Fig~\ref{fig:ari_results}).  Notably, when the agent used the Metropolis-Hastings (MH) acceptance rule, its agreement with the joint-observation posterior was significantly higher than in the Always Reject condition under both the outcome-specific and global-family Holm corrections (Table~\ref{tab:posthoc}). Its sign-based ARI was higher than in Always Reject under the outcome-specific correction, but this difference was only suggestive under the global-family sensitivity analysis ($p = 0.075$). The corresponding differences from the Always Accept condition were not statistically significant; this result does not establish equivalence between MH and AA. Human acceptance behavior also correlated with the theoretically derived MH acceptance probability (Fig~\ref{fig:human_acceptance}), suggesting that perceptual integration is enhanced by MHNG-based interaction.  Neither condition gives either agent access to the ground truth. Rather, they differ in their update rules: AA accepts the human's proposals unconditionally, whereas MH evaluates proposals against the AI's local generative model. The contribution of MH therefore lies in demonstrating an evidence-sensitive, bidirectional mechanism for constructing a shared sign system, not in demonstrating superior performance over unconditional acceptance.

A natural question is why the attained accuracy remains well below the upper bound of $\text{ARI}=1.0$. Three factors are at work, and the first is a principled limit rather than a shortcoming of the method. Each agent observes a strict subset of the generative features, so for some object pairs the union of the two observations still does not determine the ground truth partition uniquely; the joint posterior itself therefore places mass on partitions that differ from the ground truth, and the attainable ceiling is set by that posterior rather than by perfect agreement. Second, the human implements the acceptance rule only approximately, which injects noise relative to the exact sampler and slows convergence. Third, the interaction is finite, comprising $200$ naming interactions over $20$ rounds, and the sign variable is still fluctuating at the end of the session, as the trajectories in Fig.~\ref{fig:ari_results} show. Approaching the ceiling defined by the joint posterior, rather than $\text{ARI}=1.0$, is thus the appropriate target, and quantifying that ceiling directly is a worthwhile direction for future work.

It is useful to separate three levels of claim here, since they are supported to different degrees. First, what the experiment demonstrates directly: a human-AI dyad interacting under the MH acceptance rule converges on a shared sign system that agrees with the structure defined by the two agents' joint observations more closely than an agent learning from its own observations alone, and human acceptance behavior tracks the theoretically derived acceptance probability. Second, what is theoretically motivated: the MH rule implements a decentralized sampler of the posterior conditioned on the joint observations, which is why integration is expected to appear at the level of the shared sign rather than in either agent's private categories. Third, what remains open: whether a process of this kind can produce structure that is not reducible to the union of the individual observations, such as higher-order abstraction or compositionality. The present results speak to the first two levels and not to the third.

These findings empirically support MHNG theory \citep{hagiwara2019symbol, taniguchi2023emergent}, extending its applicability to human-robot systems beyond purely computational or human-human contexts \citep{okumura2023metropolishastings}. The observed alignment between human decisions and MH probabilities reinforces the idea that this rule reflects a fundamental mechanism for belief coordination and information exchange. This aligns with the CPC hypothesis \citep{taniguchi2024collective}, where the human-robot dyad operates as a decentralized Bayesian system minimizing uncertainty through local, MHNG-governed interactions. The result is a more accurate shared model of the environment.

These insights suggest a promising direction for designing AI systems capable of learning \emph{with} humans, rather than from static data or in isolation. Our dyadic interaction model can be seen as a fundamental building block of a larger human-robot ecosystem. Modeling such ecosystems, for instance with approaches from statistical physics, is crucial for understanding their macroscopic behavior, such as abrupt systemic changes depending on the human-robot composition~\citep{contucci2022human}.

We turn finally to the implications for AI alignment. The present experiment establishes a foundational result: a human and an AI agent can construct a shared sign system grounded in their complementary observations through decentralized, evidence-sensitive interaction, without either party's proposals being treated as authoritative. This provides an empirical proof of concept for the basic representational process required by co-creative learning.

The broader connection to symbiotic alignment is developed theoretically by Taniguchi et al.~\cite{taniguchi2026symbiotic}. In that framework, individual reward optimization is embedded within a broader collective free-energy objective that includes a non-decomposable collective regularization term for constructing shared representations. From this perspective, the present experiment provides initial empirical evidence that this foundational process can operate in a human-AI dyad. It does not, however, establish alignment of values or goals: our task concerns perceptual categories in a controlled laboratory game. Important challenges remain in scaling the mechanism from simple shared symbols to open-ended representations involving LLM-based agents, and in determining how symbiotic alignment can be implemented and evaluated in practical AI systems. \deleted{We therefore present co-creative learning as a candidate mechanism for bidirectional alignment \citep{shen2024towards}, rather than as evidence that such alignment has already been achieved.}\added{We therefore frame the connection to bidirectional alignment as a theoretical hypothesis, motivated by the present proof of concept and developed formally by Taniguchi et al.~\cite{taniguchi2026symbiotic}: whether co-creative learning can serve as a building block for such alignment in practical AI systems, beyond the controlled perceptual task studied here, is a direction for future work rather than a claim established by the present experiment.}

In summary, this study represents a multi-level investigation into human-robot collaboration, bridging a specific computational mechanism (MHNG) with a unifying cognitive theory (CPC/Active Inference) and validating the connection through empirical human-subject experiments. Our work underscores the value of such an interdisciplinary approach: the computational model provided a testable hypothesis about interaction, the cognitive theory offered a deep explanation for why it works, and the experimental results grounded these ideas in the reality of human-robot dynamics. This synergistic perspective is essential for making tangible progress on complex challenges like building truly symbiotic AI systems.

This study has several limitations. First, the task involved categorizing relatively simple, artificially generated visual stimuli with categorical signs. While this controlled setup was necessary, it limits the generalizability of our findings to real-world tasks involving richer semantics and more complex communication, such as those debated in the context of Large Language Models (LLMs). Second, the JA-NG paradigm assumes perfect joint attention between agents. In real-world interactions, attentional alignment can fail, making the study of mechanisms to align, control, or learn attention itself a necessary future step. Third, we do not claim that higher-order abstractions or compositional structure emerge in a ``creative'' manner. What the present experiment demonstrates is the joint construction of a shared sign system that integrates two agents' complementary partial observations. Whether a co-creative process of this kind can give rise to structure that is not reducible to the union of the individual observations is an open question, and one we regard as the natural next step for this line of work. Fourth, an agent that rejects every proposal may induce frustration or algorithm aversion, and our design cannot separate such psychological effects from the absence of information transfer. We therefore do not read the AR result as isolating information-transfer effects alone. Baselines with noisy acceptance, stochastic trust, or partially reliable agents would provide valuable and more ecologically realistic comparisons, and we regard them as important directions for future work.

This research contributes to the foundations of human-robot collaboration. Demonstrating that humans and AI can co-creatively learn and integrate perceptual information opens possibilities for AI systems that collaborate more effectively with humans, potentially enhancing human capabilities in complex decision-making, scientific discovery, or creative endeavors~\citep{cui2024collective}. The underlying theory (CPC/MHNG) offers a new perspective on AI alignment, focusing on emergent mutual understanding rather than unilateral imposition, which might lead to more robust and ethically sound AI systems in the long term. Furthermore, modeling human-robot interaction through this lens can provide insights into human communication, social learning, and cognitive science.

\section{Conclusion}
\label{sec:conclusion}

In this study, we demonstrated that a human-robot dyad can construct a shared sign system by integrating complementary partial observations through MH-governed interaction. The AI's agreement with the joint-observation posterior was significantly higher under MH than in the AR condition, and human acceptance behavior tracked the theoretically derived MH acceptance probability. Together, these findings provide empirical support for the proposed CPC mechanism in a human-robot context. This result concerns information completion under partial observability and does not imply the emergence of a qualitatively richer representation.

From the perspective of CPC, these results are consistent with a controlled instance of group-level active inference in which local interactions coordinate a shared sign system toward the posterior conditioned on the dyad's joint observations. They do not establish alignment of values or goals. Rather, they provide evidence for a candidate representational process through which future AI systems may learn with humans via mutual perceptual grounding.

Looking ahead, a significant challenge for cognitive robotics is to extend this co-creative learning model from the current abstract task to physically embodied agents engaged in shared physical tasks with humans. From the perspective of symbol emergence in robotics, a crucial next step is to investigate whether the underlying mechanism can scale to the formation of more complex, compositional symbol systems, enabling robots to communicate about objects, actions, and their relationships in a more nuanced way. Overcoming these challenges will be key to realizing the vision of truly symbiotic human-robot systems that this work helps to establish.

\section*{Data Availability}
The datasets, code, and other digital research materials supporting the results of this study are available via Zenodo at \url{https://doi.org/10.5281/zenodo.15873161}~\citep{zenodo}.

\section*{Acknowledgement(s)}

We gratefully acknowledge the participants for their contributions to the experiment.

\section*{Funding}

This work was supported by JSPS KAKENHI Grant Numbers JP23H04835 and JP21H04904, as well as the JST Moonshot R\&D Grant Number JPMJMS2011.

\bibliographystyle{tfnlm} 
\bibliography{co-creative}

\newpage

\appendix
\section{Ethics Statement}
\label{sec:appendix_ethics}
The research project, titled "Co-creative learning in human-machine systems based on joint-attention naming game," received approval from the Research Ethics Committee for Non-Medical Life Science Research at Ritsumeikan University (Approval No. BKC-LSMH-2023-095). The approved research period extends until March 31, 2026. The study was classified as non-invasive research without intervention and underwent rapid review. All 90 participants were recruited online via the crowdsourcing platform CrowdWorks. Informed consent was obtained digitally through the experimental web application prior to participation, after participants reviewed the information sheet detailing the study's purpose, procedure, potential risks (minimal, primarily eye strain from screen time), data handling, compensation (2000 JPY), and voluntary nature (right to withdraw anytime without penalty). Data collected included CrowdWorks IDs (for payment, immediately replaced with anonymized research IDs after confirmation), basic demographics (age, gender), and experimental data (image classifications, names proposed, acceptance/rejection decisions, optional free-text feedback).

\section{Instructions Provided to Participants}

\begin{figure}[tb!p]
  \centering
  \includegraphics[width=0.9\linewidth]{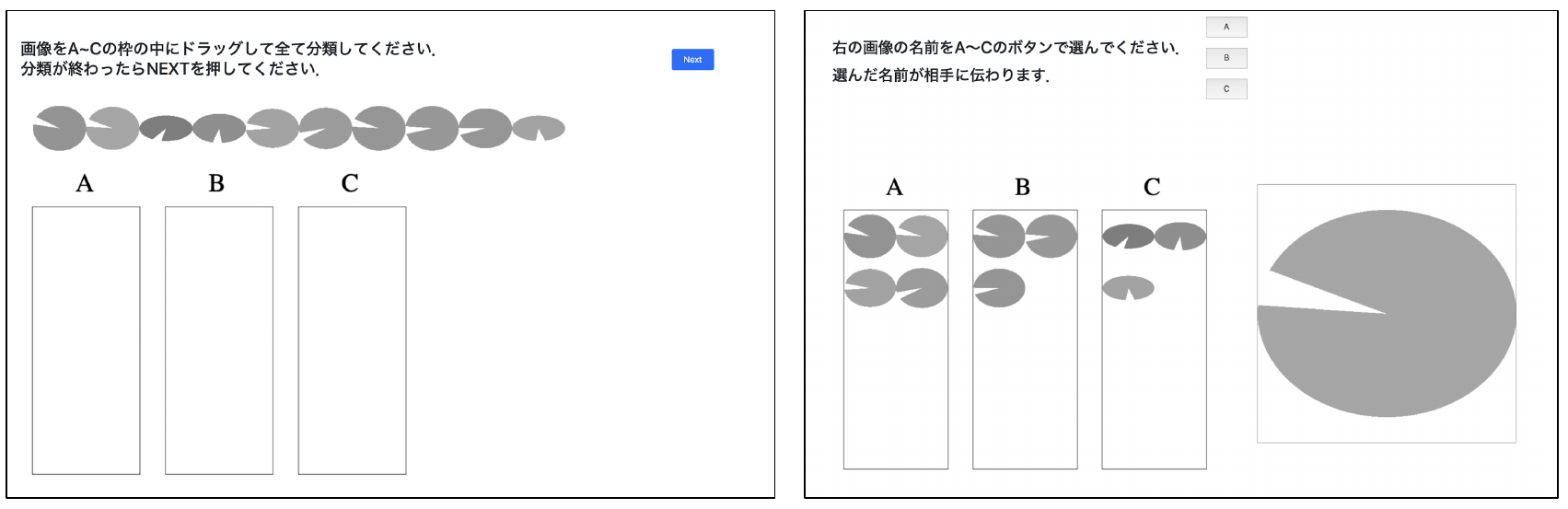}
  \caption{Screenshot of the experimental interface used in the JA-NG task. The graphical user interface supports partial observation and sign exchange between human and computer agents.}
  \label{fig:experiment_screenshot}
\end{figure}

After being recruited via CrowdWorks, participants were directed to the web-based experimental system. There, they were presented with a series of written instructions including purpose and methodology. The main parts of the methodology are shown below.
To aid understanding, a diagram illustrating the experimental process was also provided.
As all participants in the experiment were native Japanese speakers, the instructions were originally provided in Japanese. Below, we provide English translations of the key parts of the instructions on methodology. Figure~\ref{fig:experiment_screenshot} shows a screenshot of the experimental interface.

\begin{tcolorbox}[breakable, enhanced, colback=white, colframe=black, title=Instructions to participants]

\subsection*{Methodology}

\subsubsection*{Study Design}

This experiment is conducted online using a web-based application. Participants use their own personal computer (PC) and carry out the experiment within a browser-based interface.

\subsubsection*{Procedure and Duration}

\begin{enumerate}
    \item \textbf{Explanation and Consent} (current stage): You are now reading the explanation of the experiment. If you agree to participate, we will obtain your informed consent via the web interface. If you do not agree, the experiment will terminate immediately. (Estimated time: 20 minutes)\\
    If you have any questions about the study, please contact us at: [email address]
    
    \item \textbf{Communication Experiment using the Web Application} (90 minutes)
    
    \item \textbf{Post-Experiment Questionnaire} (10 minutes)
    
    \item \textbf{End of Experiment}
\end{enumerate}

If the experiment page is accidentally closed due to a system error or user action, it can be resumed by refreshing or reopening the link. You may freely pause or withdraw at any point. In the event that technical issues prevent resumption, please contact us via the email address above.

\subsection*{Overview of the Experimental Procedure}

The human and the computer each observe different parts of the same images. Because each has incomplete information, communication through a naming game is essential for improving classification accuracy.

\paragraph*{Step-by-Step Procedure:}

\begin{enumerate}
    \item \textbf{Initial Categorization:} Classify 10 images into three categories labeled A–C.
    
    \item \textbf{Proposal and Response:} The participant and the computer take turns as the “namer” and the “listener.” The namer selects one of the images and proposes a label (A–C). The listener chooses to accept or reject the label by clicking a button.
    
    \item \textbf{Label Adjustment:} If the listener accepts the label and needs to revise their classification or labeling method, they may change their categorization freely.
    
    \item \textbf{Role Switching:} The roles are reversed and steps (2) and (3) are repeated.
    
    \item \textbf{Cycle:} Steps (2)–(4) are conducted for each of the 10 images.
    
    \item \textbf{Repetition:} The above cycle is repeated for a total of 20 rounds.
\end{enumerate}

\paragraph*{Rules for Label Acceptance and Category Update:}

\begin{itemize}
    \item \textbf{If both you and the computer have the same label:}\\
    Accepting the proposal $\rightarrow$ No change is needed.
    
    \item \textbf{If you and the computer have different labels:}
    \begin{itemize}
        \item Accepting the proposal $\rightarrow$ Change your categorization to match the proposed label.
        \item Rejecting the proposal $\rightarrow$ Keep your current categorization unchanged.
    \end{itemize}
\end{itemize}

Your objective is to collaborate with the computer agent to improve the accuracy of classification. As noted, both you and the computer are presented with only partial information. Thus, effective communication in the naming game is essential for sharing the missing information and reaching mutual understanding.
An example of the application interface used in the experiment is shown in the following figures.

\end{tcolorbox}

\end{document}